\documentclass[reprint,amsmath,amssymb,aps,pra]{revtex4-1}
\usepackage{graphicx}% Include figure files
\usepackage{dcolumn}% Align table columns on decimal point
\usepackage{bm}% bold math
\usepackage{hyperref}% add hypertext capabilities
\usepackage[mathlines]{lineno}% Enable numbering of text and display math
\usepackage{eurosym}

\begin{document}

\title{Effective $p$-wave Fermi-Fermi Interaction Induced by Bosonic Superfluids}% Force line breaks with \\

\author{Yongzheng Wu}
\email{yzwu15@fudan.edu.cn}
\affiliation{Department of Physics and State Key Laboratory of Surface Physics, Fudan University, Shanghai 200438, China}
\author{Zheng Yan}
\affiliation{Department of Physics and State Key Laboratory of Surface Physics, Fudan University, Shanghai 200438, China}
\author{Zhi Lin}
\affiliation{Department of Physics and State Key Laboratory of Surface Physics, Fudan University, Shanghai 200438, China}
\author{Jie Lou}
\thanks{Corresponding author}
\email{loujie@fudan.edu.cn}
\affiliation{Department of Physics and State Key Laboratory of Surface Physics, Fudan University, Shanghai 200438, China}
%\affiliation{Collaborative Innovation Center of Advanced Microstructures, Nanjing 210093, China}
\author{Yan Chen}
\thanks{Corresponding author}
\email{yanchen99@fudan.edu.cn}
\affiliation{Department of Physics and State Key Laboratory of Surface Physics, Fudan University, Shanghai 200438, China}
\date{\today}%

\begin{abstract}
We study the two-dimensional Bose-Fermi mixture on square lattice at finite temperature by using the determinant quantum Monte Carlo method within the weakly interacting regime. Here we consider the attractive Bose-Hubbard model and free spinless fermions. In the absence of boson-fermion interactions, we obtain the boundary of the collapsed state of the attractive bosons. In the presence of boson-fermion interactions, an effective $p$-wave interaction between fermions will be induced as far as the bosons are in a superfluid state. Moreover, we find the emergence of the composite fermion pairs at low temperatures.
\end{abstract}

\maketitle

%\tableofcontents

\section{\label{sec:level1}Introduction}

During the past years, the techniques of loading ultra-cold atoms gases on the optical lattices have been extensively explored in simulating quantum many-body systems. Various lattice models with strong correlations, such as the Bose or Fermi-Hubbard model\cite{Sci-OcsFhm2016}, have been realized experimentally by using the optical lattices.  Besides, with the Feshbach resonances technique \cite{S.Inouye1998, S.L.Cornish2000, RMP-Frug2010}, interactions between particles can be tuned from attractive to repulsive in a controlled way, which is not entirely possible in other systems like the solid state materials. The advancements in the experimental techniques have made it possible to study fundamental and interesting many-body physics, for example, the superfluid to Mott insulator transition \cite{Nat-qptms2002} of dilute Bose gas, the Bardeen-Cooper-Schrieffer (BCS) to Bose-Einstein Condensate (BEC) crossover in the Fermi gas \cite{PRL-Orcfap2004}. Another landmark achievement which exhibits the prominent advantages of optical lattice systems is the observation of maximal antiferromagnetic spin correlations in the 2D Fermi-Hubbard model at half-filling \cite{Sci-OcsFhm2016}. Furthermore, the mixtures of Bose and Fermi superfluid have been realized by using dilute gases of two lithium isotopes, ${}^{6}Li$ and ${}^{7}Li$ \cite{Sci-Mbfs2014}, as well as two different alkali elements, ${}^{174}Yb$ and ${}^{6}Li$ \cite{Richard-Roy2017}. The experimental progress has stimulated lots of interest in the theoretical study of Bose-Fermi mixtures at low temperature recently.

On the other hand, it has been shown that in a 3D optical lattice with dilute Bose-Fermi mixture, interactions between the two-component fermions can be induced due to the density fluctuations of superfluid bosons \cite{Pra-pfb2000}. Two fermionic atoms in such systems can interact with each other by exchanging a phonon that propagates through the Bose condensate, which acts a role similar to the phonons in the BCS superconductivity. The emergence of topological $p+ip$ fermionic superfluid with a high critical temperature was recently proposed for a 2D spin-polarized Fermi gas immersed in a 3D BEC \cite{Prl-tpsfb2016}. In particular, the fermions attract each other via an induced
interaction mediated by the bosons.  Moreover, previous numerical studies suggest that the composite fermionic pairs (particle-particle or particle-hole), which are formed by the bosons and fermions in the mixtures, show standard Fermi liquid or polaronic behaviors \cite{Lode_pollet2006, Pra-bfpo2008, Pra-Bfplt2013}. However, most of the previous studies have been focused on 1D or mixed dimensions (2D-3D) or 3D systems.

In this paper, we investigate the 2D Bose-Fermi mixture system on the square lattice by using the determinant quantum Monte Carlo (DQMC) method in the parameter regimes free of sign problem. Our calculations obtain the boundary of collapse state of the attractive Bose-Hubbard model at finite boson density. The numerical results reveal the emergence of an effective $p$-wave interaction between two fermions, which is induced by the underlying bosonic superfluid. Moreover, we find the appearance of the composite fermion pairs at low temperatures.

Here we consider that a homogeneous mixture of ultracold bosons and spinless fermions is loaded into an optical lattice with a square well potential. The system can be described by the Bose-Fermi Hubbard model. Here we consider a mixture composed of single component bosons and spinless fermions. The corresponding Hamiltonian for such a system is given by:
\begin{equation}\label{1}
\begin{aligned}
&{H_{bf}}=-\sum\limits_{\left\langle {ij} \right\rangle } {({t_b}b_i^\dag {b_j}+ {t_f} c_{i}^\dag {c_{j}} + H.c.)}   + \sum\limits_i {{\varepsilon _b}{n_i}}  + \sum\limits_i {{\varepsilon _f}{m_i}} {\kern 1pt} \\
& {\kern 1pt} {\kern 1pt} {\kern 1pt} {\kern 1pt} {\kern 1pt} {\kern 1pt} {\kern 1pt} {\kern 1pt} {\kern 1pt} {\kern 1pt} {\kern 1pt} {\kern 1pt} {\kern 1pt} {\kern 1pt} {\kern 1pt} {\kern 1pt} {\kern 1pt} {\kern 1pt} {\kern 1pt} {\kern 1pt} {\kern 1pt} {\kern 1pt} {\kern 1pt} {\kern 1pt} {\kern 1pt} {\kern 1pt} {\kern 1pt} {\kern 1pt} {\kern 1pt}  + \frac{{{U_b}}}{2}\sum\limits_i {n_i^2} + {U_{bf}}\sum\limits_i {{n_i}{m_i}}
\end{aligned}
\end{equation}
where the operator $b_i^\dag$ ($b_i$) creates (annihilates) a boson on site $i$ while $c_{i}^\dag$ and $c_{i}$ are the corresponding fermionic operators, $n_i=b_i^\dag b_i$ $(m_i=c_{i}^\dag c_{i})$ corresponds to the boson (fermion) number operator, $t_b$ and $t_f$ are the bosonic and fermionic hopping integrals between two nearest neighboring sites, $\varepsilon _b$ $(\varepsilon _f)$ represents the chemical potential for bosons (fermions), $U_b$ is the two bodies interaction between bosons, and $U_{bf}$ is the coupling strength between bosons and fermions. Throughout this paper, we take $t_b=t_f=1$ as the energy unit.

We employ the finite temperature determinant Monte Carlo (DQMC) method \cite{R-Blankenbecler-1981, D-J-Scalapino-1981, J.E.Hirsch1982, J.E.Hirsch1985, Raimundo-R-dos-Santos-2003, Wirawan-Purwanto-2004} for numerical simulations.  It is a field-theoretic method where many-body propagators resulting from two-body interactions can be transformed into one-body propagators by using the Hubbard-Stratonovich (HS) transformation. The resulting integrals can then be computed by Monte Carlo sampling. In Ref. \cite{Brenda-M-Rubenstein-2012}, the DQMC method was first used to study Bose-Fermi mixtures via a combination of bosonic and fermionic Monte Carlo techniques. Though some exact results can be obtained in the small size lattices, the sign problem is still quite severe in most parameter ranges (strong repulsive interaction in the large size lattices at low temperature). Interestingly, we find that there are two cases where the sign of the determinant is always equal to one. One example is the attractive Bose-Hubbard model at low particle density, whose Hamiltonian can be acquired by ignoring the whole fermionic part of the Eq. (1). Another case is in the Bose-Fermi mixture with the interaction between bosons and fermions being weaker than that of the bosons, namely $|U_{bf}| \le |U_b|$ and $U_b \le 0$. On the other hand, a repulsion between bosons and fermions prefers a demixing to minimize the overlapping region, whereas in the case of an attraction the mixture can collapse, as long as the particle numbers are sufficiently large or interspecies interaction is strong \cite{C.Ospelkaus2006, G.Modugno2002}. In the following, we mainly focus on dilute Bose-Fermi mixture with weak interspecies interaction where the sign of the corresponding determinants in these two cases which are used to express the trace over one-body propagators is always positive in our calculation (Monte Carlo samples up to $10^4$) so that the many-dimensional integrals can be performed accurately by the Monte Carlo sampling.

\section{\label{sec:level1}Bose-Hubbard model with attractive interaction}
In this section, we investigate the Bose-Hubbard model by considering only the bosonic part of the Hamiltonian in Eq. (1). According to the previous studies on the atomic Bose gas with negative scattering lengths, the number of the bosons would be quite limited due to the collapse of the BEC \cite{H-T-C-Stoof-1994, C-C-Bradley-1996, J-L-Roberts-2001}. More specifically, the maximum of the boson number $N_c$ of a system under experiment is inversely proportional to the negative scattering length $a_s$, that is $N_c \cdot a_s = constant$. Here we choose the density $n_b$ of a square lattice system with size $6\times6$ to be $0.02\pm0.0006$ which could be realized within experimental reach \cite{Pra-pfb2000}. The on-site attractive interactions are set to be $U_b=-0.1, -0.2, -0.3$.

\subsection{\label{sec:level2} Off-diagonal long-range order of bosons}
Here we study the superfluidity of bosons at low temperature. The off-diagonal correlation function $\bar g(r) =\langle  b_i b_j^\dagger \rangle $ \cite{O-Penrose1956,Anthony-J-Leggett2001}, where $r=|i-j|$ is the distance between two lattice site $i$ and $j$, has been investigated. We rewrite the correlation function as $g(r) =\langle   b_i^\dagger b_j \rangle $ for a better comparison.  The calculations are performed for several different inverse temperatures with lattice size $4\times4$, $6\times6$, $8\times8$, and the corresponding results are presented in Fig. 1(a), (b), and (c) respectively. Note that the bosons in our system are quite dilute. As a result, the on-site interactions have little effects on the density distribution or the off-diagonal correlation, and we only present results with $U_b=-0.2$. From the off-diagonal correlation function shown in Fig. 1(a), (b) and (c), we can observe that at the high temperature such as $\beta=0.5$, the off-diagonal correlation goes down to zero quickly. While as the temperature is progressively reduced, the decay of the correlation become more and more slowly with a finite value at the longest distance $r=r_{max}$ which indicates that the finite size system has entered the superfluid state.

To show finite size effect, finite size extrapolations of the off-diagonal long-range order $g(r_{max})$ has been done for lattice size $4\times4$, $6\times6$, $8\times8$ at various inverse temperature. The results are shown in Fig. 1(d). For high temperature $\beta=0.5, 1.0, 2.0$, the off-diagonal long-range order decay quickly to zero. While a finite value $g(r_{max}) \approx 0.014$ can be obtained by doing linear fitting at $\beta = 5.0$. To verify the superfluidity of bosons in the thermodynamic limit, larger system sizes are need to be considered.

%We also show the off-diagonal correlation in Fig. 1(d). It is clear that at a lower temperature, the correlation may decay slower. At a relative high temperature, $\beta = 0.5$, the correlation may decay to zero around $r=3$. However, for $\beta = 2.0$, $\bar{g}(r)$ remains to be a finite value at the longest distance, which means that the system persists the superfluid state.

\begin{figure}
  \centering
  \includegraphics[width=9.0cm]{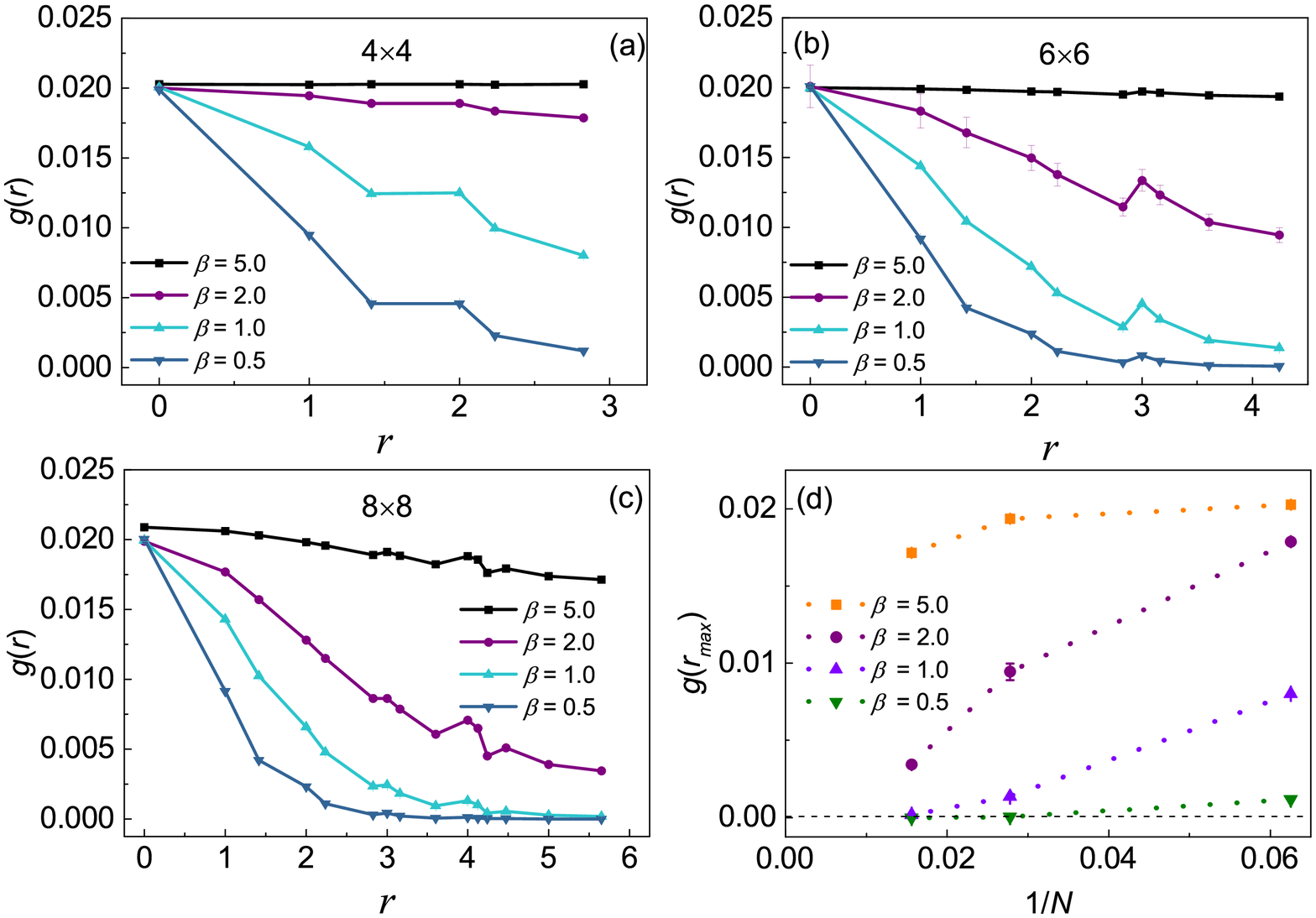}
  \caption{The off-diagonal correlation function $g(r)$ of the boson at different inverse temperatures for lattice sizes $4\times4$ (a), $6\times6$ (b) and $8\times8$ (c). The finite size extrapolations have been done for $g(r_{max})$ to show the size effect (d). Here $N =4\times4, 6\times6, 8\times8$ and the interaction is set to be $U_b=-0.2$.}\label{1}
\end{figure}

\subsection{\label{sec:level3} Critical value of the collapsed state at finite temperature}
The influence of a larger number of bosons, which can be tuned by increasing the chemical potential $\varepsilon_b$, can also be analyzed by the numerical method. Interestingly, the particle densities at certain lattice sites increase abruptly while densities at other sites are decreased. This phenomenon is consistent with the collapsed state observed in previous experimental studies \cite{J-L-Roberts-2001}. To better characterize the collapsed state, we introduce the inverse participation ratio (IPR) $\alpha$ \cite{Dean-bell-1970}, which is usually adopted as the criteria for localization in numerical studies. The IPR is defined as :
\begin{equation}\label{2}
 \alpha  = \sum\limits_i^N {{{\left| {{\alpha _i}} \right|}^4}} /{\left( {\sum\limits_i^N {{{\left| \alpha_i  \right|}^2}} } \right)^2}
\end{equation}
where $\alpha_i$ is the particle density at site $i$, and $N$ is the number of lattice sites. $\alpha$ has a small value of $O(N^{-1})$ when the system is in a uniform state. However, it will reach 1 when all the particles are localized at a certain lattice site. In the inset of Fig. 2, the variation of $\alpha$ as the chemical potential $\epsilon_b$ changed is presented at the inverse temperature of $\beta=0.5$. It can be found that when the system is in the normal state, the IPR is a constant value $\alpha = 1/36$. While if the system enters the collapsed state in which the system becomes unstable, the IPR will suddenly jump to a much larger value closed to 1. For example, when $U_b=-0.3$, as indicated by the blue dotted line in the inset, $\alpha$ is quite close to 1. As we suppress the attractive interaction between the bosons, the abrupt change of IPR still shows up as the system goes into the collapsed state (see the dotted lines corresponding to $U_b=-0.2$ and $-0.1$, respectively). We have also calculated the IPR of the system at several other temperatures (not shown here), and find that when the system undergoes a phase transition between a normal state and the collapsed state, the IPR will always show a sudden jump at the critical value.

In Fig. 2, we also show the critical value of the boson density $n_c$ which corresponds to the onset of the sudden jump of IPR for different temperatures $\beta=0.25$, 0.5 and 3.0. It can be observed that the critical value will be reduced as we enhance the attractive interaction between the bosons. Besides, as the temperature is lowered down (i.e., with $\beta$ increasing from 0.25 to 3.0), the reduction of the critical value will become sharper. This is consistent with the result from the previous study in which the equation describes the connection between the critical value and the interaction strength $n_c \times |U_b|=k$, with $k$ being a constant which decreases for lower temperatures. A lattice of size $4 \times 4$ has also been investigated, and similar results are obtained (results not shown here). We can conclude that the regime of the collapsed state will be enlarged as the temperature goes down.
\begin{figure}
  \centering
  \includegraphics[width=8.4cm]{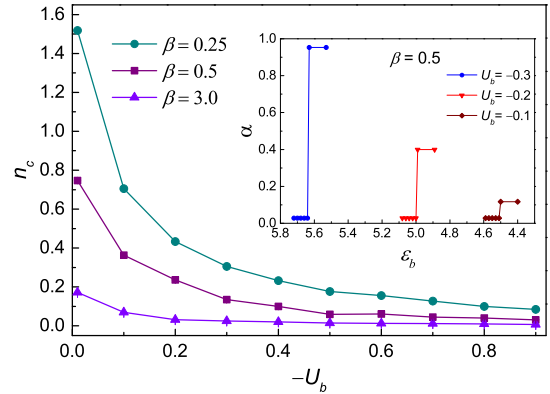}\\
  \caption{The critical value of density vs the absolute value of attractive interaction $U_b$ for various temperatures with lattice size $6\times 6$. Obviously, the region of the collapse state enlarges as the temperature goes down. The inset displays the participation ratio of density versus the chemical potential $\varepsilon_b$ for several attractive interaction at inverse temperature $\beta=0.5$. Note that normally the density of boson increases linearly as the the chemical potential decreases.}\label{3}
\end{figure}

\section{\label{sec:level1}Bose-Fermi mixture with spinless fermions}
\subsection{\label{sec:level2} Setup}
%After checking the results for the attractive Bose-Hubbard model with our numerical method, now we can move forward to deal with the Bose-Fermi mixture systems.
For the sake of simplicity, we consider the spinless fermion only in the Bose-Fermi mixture system. To avoid the sign problem, we set the interaction strength between the bosons and fermions $U_{bf}$ to be equal to the interaction between bosons $U_b$ unless otherwise stated. Here we mainly concentrate on the square lattice system with size $N = 6 \times 6$ in which densities of bosons and fermions are tuned to be $\langle n_b \rangle=0.02\pm0.0006$ and $\langle m_f \rangle=0.2\pm 0.007$ respectively. To show finite size effect of $p$-wave pair correlation function, different lattice sizes $4\times4$ and $8\times8$ have also been investigated. Here we choose a relatively low temperature with $\beta = 5.0$. Besides, to avoid the collapse of the superfluid state in the attractive Bose system, which is shown in Sec. II, the interactions in the mixtures should be chosen to be very weak at low temperature. So we set $-|U_{bf}| = U_b = -0.05$, $-0.1$ and $-0.2$. It is noteworthy that the boson-fermion interaction will change from $U_{bf}$ to $-|U_{bf}|$ under particle-hole transformation, which can be shown by changing the operator $c_i$ to $(-1)^i c_i^\dagger$ in Eq. (1) \cite{Lode_pollet2006}.

\subsection{\label{sec:level2} Charge fluctuation and $p$-wave pair correlation}
Due to the coherence of the superfluid bosons and interaction between boson and fermion, two fermions can affect each other via the bosons in the mixture. To study the influences of the superfluidity on the fermions, we calculate the equal time charge fluctuation of the fermions and the density fluctuation between bosons and fermions. The numerical results for systems with various coupling strengths are presented in Fig. 3. Generally speaking, the charge fluctuation of fermions is given by $C(r)=\langle m_i m_j \rangle -\langle m_i\rangle \langle m_j \rangle$ with $r=|i-j|$. Note that the angle brackets denotes both the quantum mechanical average over the whole space-time configuration and the statistical average of the samples (HS field configurations)\cite{Raimundo-R-dos-Santos-2003}.  To reduce the statistical errors, we redefine charge fluctuation as:
\begin{multline}\
\begin{array}{l}
\bar{C}(r)
 = \left\langle {\tilde{g_f}(i,i) \tilde{g_f}(j,j)} \right\rangle  +  \left\langle {\tilde{g_f}(j,i) {g_f}(i,j)}   \right\rangle \\
{\kern 1pt}  - \left\langle {\tilde{g_f}(i,i)} \right\rangle \left\langle {\tilde{g_f}(j,j)} \right\rangle  - \left\langle {\tilde{g_f}(j,i)} \right\rangle \left\langle {{g_f}(i,j)} \right\rangle
\end{array}
\end{multline}
where the Wick¡¯s theorem have been used to decompose the time ordered expectation value of two-body operators into the expectation value of the single fermionic Green¡¯s function $g_f(i,j) = c_i^\dagger c_j$ and $\tilde{g_f}(i,j)=\delta_{i,j}-g_f(i,j)$. The last term in Eq. (3) is introduced because each term like $\langle g_f g_f \rangle$ corresponds to a term of the form $\langle g_f \rangle \langle g_f \rangle$. Under this definition, the charge fluctuation will be always strictly equal to zero as long as there is no coupling between the bosons and fermions. Similarly, the density fluctuation between bosons and fermions is defined as :
\begin{equation}\label{4}
  \bar{C}_{bf}(r) = \left\langle  {{n_i}{m_{i + r}}} \right\rangle  - \left\langle {{n_i}}  \right\rangle \left\langle {{m_{i + r}}} \right\rangle
\end{equation}
By applying the particle-hole transformation to Eq. (3) and Eq. (4), it is easy to check that $\bar{C} (r)$ may remain unchanged while $\bar{C}_{bf}(r)$ may change its sign. From Fig. 3, we can find that the charge fluctuation is enhanced at the short-range due to the occurrence of the superfluid bosons. The sign of the interaction between the bosons and fermions has no evident effect on the results within numerical errors. In addition, a long range correlation between the bosons and fermions can be observed from the $\bar{C}_{bf}(r)$, as shown in the inset of Fig. 3. Here for positive and negative $U_{bf}$, the density fluctuations are symmetric about line of zero fluctuation. This is consistent with our analysis.

\begin{figure}
  \centering
  \includegraphics[width=8.4cm]{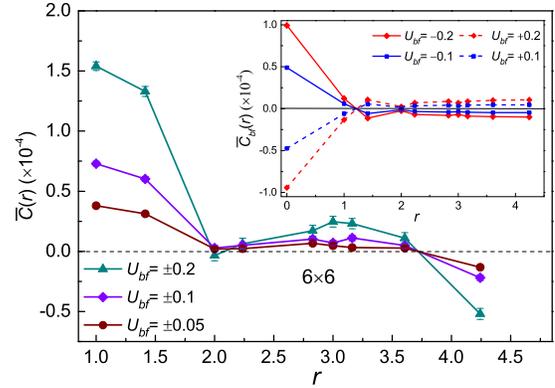}\\
  \caption{Charge fluctuation versus distance $r$ with Bose-Fermi couplings $U_{bf}=\pm0.05,\pm0.1,\pm0.2$ at inverse temperature $\beta=5.0$. The inset is the density fluctuation between boson and fermion with various couplings. The density fluctuation remain to be a finite value at long range denotes that the long range off-diagonal order may exist between boson and fermion.}\label{5}
\end{figure}

More interestingly, due to the interaction between the bosonic and fermionic components in the Bose-Fermi mixture, the effective $p$-wave pairing interactions between fermions could be induced. The $p$-wave pair correlation is defined as $P(r) = \langle \Delta^\dagger (m) \Delta (n) \rangle$. Here $\Delta (m) = \sum_{\vec{\sigma}} f(\vec{\sigma}) c_{\vec{m}} c_{\vec{m}+\vec{\sigma}}$ with $f(\vec{\sigma})=0$ for $\vec{\sigma}=\pm \vec{y}$ and $f(\vec{\sigma})=\pm 1$ for $\vec{\sigma}=\pm \vec{x}$ respectively. $m$ ($n$) corresponds to the lattice site involving the $p$-wave pairing. Similar to the definition of the charge fluctuation, we can also redefine the $p$-wave pair correlation as
\begin{equation}\
  \bar{P}(r) = P\left( r \right) - P{\left( r \right)_0}
\end{equation}
where $P(r)_0$ is the uncorrelated pair correlation \cite{Shiwei-Zhang1997,Z.B.Huang2001}. Namely, for each $\langle  g_f g_f \rangle$ in $P(r)$, there will be a term of the form $\langle g_f  \rangle \langle g_f \rangle$. We have calculated the pair correlation along both the x and y-direction in our numerical calculations and the results are shown in Fig. 4. $\bar{P}(r)$ becomes weaker as the distance increases in both directions. The sign of $U_{bf}$ has no significant effect on the results. From Fig. 4(a), we can also find that at the same distance, a stronger coupling between bosons and fermions will lead to a stronger $p$-wave pairing and the sign of $\bar{P}(r)$ will not change. However, in the x-direction (see Fig. 4(b)), the correlation will change sign at the same distance while the strength of $U_{bf}$ changes. The conclusion that stronger coupling between bosons and fermions will result in stronger $p$-wave pair correlation still holds along the x-direction. So the pair correlation will always be zero if there is no coupling between the bosons and fermions but will become finite if $U_{bf}$ is turned on. We can conclude that the superfluid bosons mediate the effective $p$-wave pairing between the fermions.
\begin{figure}
  \centering
  \includegraphics[width=8.4cm]{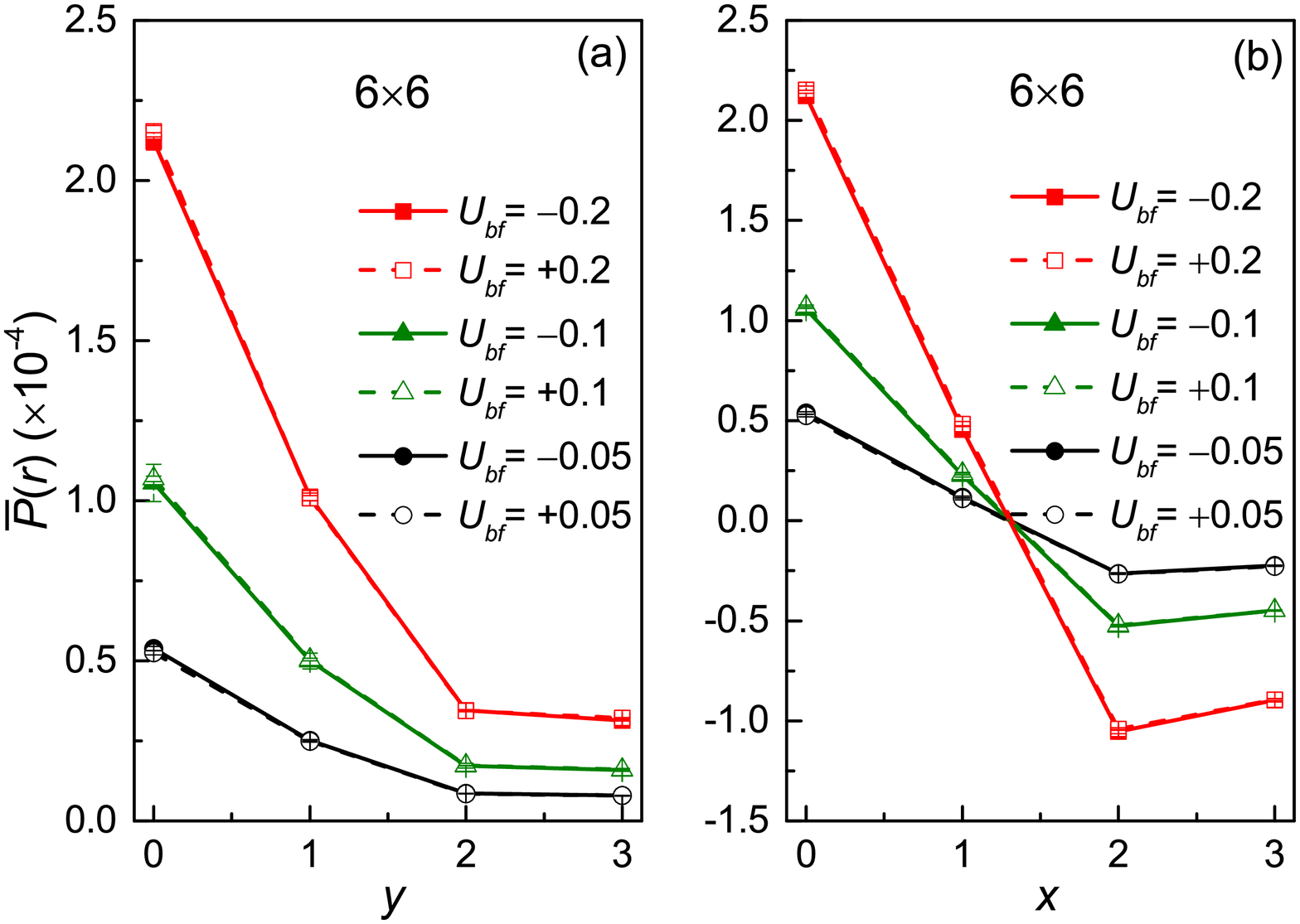}\\
  \caption{$P$-wave pair correlation with Bose-Fermi interactions $U_{bf}=\pm0.05,\pm0.1,\pm0.2$ at inverse temperature $\beta=5.0$. (a) along y-direction of lattice, (b) along x-direction of lattice.}\label{5}
\end{figure}

At this stage, we need to address whether the superfluidity of the boson is a necessary condition for the induction of $p$-wave correlation of fermions. According to the results shown in Fig. 1, the bosons enter the superfluid state, which is demonstrated to be the second order phase transitions, with the temperature goes down. This superfluid state often can be described by the off-diagonal long-range order. Here we consider the $g(r_{max})$ as the long-range order parameter of the bosons with $r_{max}=3 \sqrt 2$ for system sites $6\times6$. Meanwhile, we take the $p$-wave correlation $\bar{P}(r_{max})$ as the off-diagonal long-range order parameter for the $p$-wave of the fermion with $r_{max}=3$ in the y-direction. For clarification, $\bar{P}(r_{max})$ of the different interaction strengths $U_{bf}$ are rescaled to the $U_{bf}=-0.1$ by dividing $|U_{bf}|$. And $g(r_{max})$ is divided by 1200. These results are shown in Fig. 5. Note that the weak Bose-Fermi couplings have little effect on $g(r_{max})$, so we only show the $g(r)$ of $U_{bf} = -0.2$ in the figure. It is obvious to find that the boson goes into the superfluid state as the temperature goes down shown by the real olive line. Due to the finite size effect, $g(r_{max})/1200$ turns into zero smoothly. It is important that the $\bar{P}(r_{max})$ of the different interaction strengths $U_{bf}$ (dashed line) show the almost same trend with $g(r_{max})$, which indicate that the superfluid state of the bosons is a prerequisite for the $p$-wave correlation of the fermions.

\begin{figure}
  \centering
  \includegraphics[width=8.4cm]{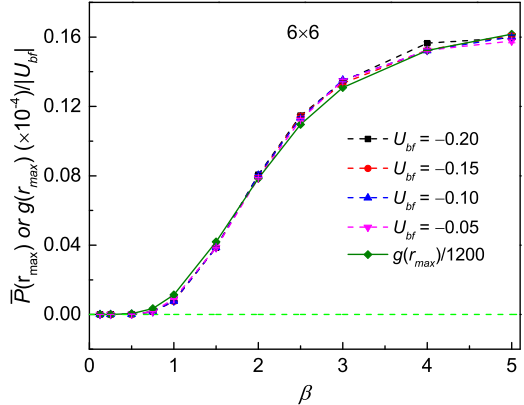}\\
  \caption{$P$-wave pair correlation $\bar{P}(r_{max})$ along y-direction of the lattice for interactions $U_{bf}=-0.05,-0.10,-0.15,-0.20$ at various inverse temperature $\beta$ and the off-diagonal long range order $g(r_{max})$ of the boson at various temperatures. $\bar{P}(r_{max})$ are resacled by dividing $|U_{bf}|$ and $g(r_{max})$ is rescaled by dividing 1200. $\bar{P}(r_{max})$ and $g(r_{max})$ show almost same trend.}\label{Fig5}
\end{figure}

Both the charge fluctuation and the $p$-wave pair correlation indicate that an effective interaction \cite{M.Lewenstein2004} can be induced by the superfluid bosons. To further understand this effective interaction, we use the DQMC method to calculate a simplified spinless fermion model whose Hamiltonian is given by :
\begin{equation}\label{}
  {H_f} =  - {t_f}\sum\limits_{\left\langle {ij} \right\rangle } {c_{i}^\dag {c_{j}}} {\kern 1pt} {\rm{ + }}{{\rm{V}}_{eff}}\sum\limits_{\left\langle i j \right\rangle} {{m_i}{m_{j}}} {\kern 1pt} {\kern 1pt}  + \sum\limits_i {{\varepsilon _f}{m_i}}
\end{equation}
The $V_{eff}$ here represents the effective interaction between the nearest neighboring fermions. Since the boson-fermion coupling is weak, we can neglect the interactions between fermions separated by larger distances and only consider the nearest-neighboring situation. Generally, the spinless fermion model is sensitive to the sign problem at low temperature except the positive definite case \cite{Zi-Xiang-Li2015}. Fortunately, we just care about the weak interaction region $(V_{eff}\in[-0.5,0])$ at the low density $(\left\langle m \right\rangle =0.2)$ where the sign is always positive with samples up to $10^4$. For a better comparison, we introduce the ratio $\gamma_p (r) = \bar{P}(r)_{bf}/\bar{P}(r)_{eff}$ where $\bar{P}(r)_{bf}$ and $\bar{P}(r)_{eff}$ are the $p$-wave pair correlation at distance $r$ in the mixture and spinless fermion system respectively. It is similar to the ratio of charge fluctuation $\gamma_c(r)=\bar{C}(r)_{bf}/\bar{C}(r)_{eff}$. The numerical results about the $\bar{C}(r)$ and $\bar{P}(r)$ for lattice sizes $6\times6$ and $8\times8$ are shown in Fig. 6. Surprisingly, both the ratios of charge fluctuation and $p$-wave pair correlation $\gamma(r)_{p/c}$ at $|U_{bf}|=0.05, 0.1, 0.2$ are closed to 1.0($\pm 0.1$). The corresponding effective attractive interaction $-V_{eff}=0.025, 0.05, 0.1$, respectively. The results are shown in Fig. 6 (violet line). These results indicate that an effective interaction between fermions are indeed induced by the superfluid bosons. What's more, the strength of effective interaction is a half of the Bose-Fermi interaction, which is much different form the induced effective interaction in 3D Bose-Fermi mixture\cite{Pra-pfb2000}.

\begin{figure}
  \centering
  \includegraphics[width=9.0cm]{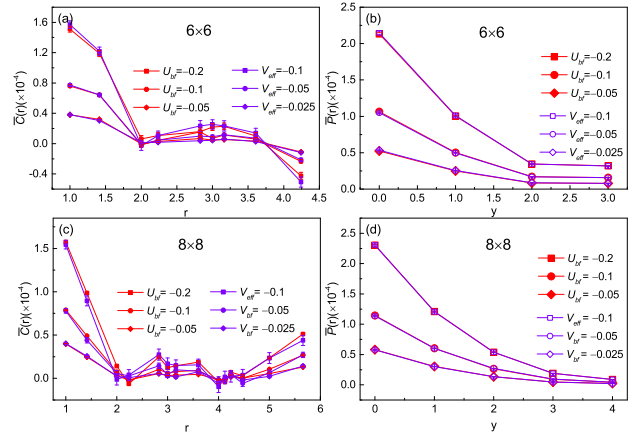}\\
  \caption{Charge fluctuation (a) and $p$-wave pair correlation along y-direction (b) for Bose-Fermi interactions $U_{bf}=-0.2,-0.1,-0.05$ (red line) and $V_{eff}=-0.1,-0.05,-0.025$ (violet line) respectively at same inverse temperature $\beta=5.0$ with lattice size $6\times6$. Here (c) and (d) are corresponding results of the charge fluctuation and $p$-wave pair correlation for lattice size $8\times8$.}\label{5}
\end{figure}

\subsection{\label{sec:level2} Finite size extrapolations for P-wave pair correlation function}
As the temperature goes down, the $p$-wave pair correlation decays slowly shown in the inset of Fig. 7. It is obvious that the lower of the temperature, the slower decay of the $p$-wave pair correlation. And a finite value at the longest distance $r=r_{max}=3$, which is served as the off-diagonal long range order of $p$-wave pair correlation function, can be observed at low temperature $\beta=2.0, 5.0$ for lattice size $6\times6$.

To investigate the finite size effect, the $p$-wave pair correlation functions for different lattice sizes $4\times4$, $6\times6$, $8\times8$ have been calculated at various inverse temperatures.
\begin{figure}
  \centering
  \includegraphics[width=8.4cm]{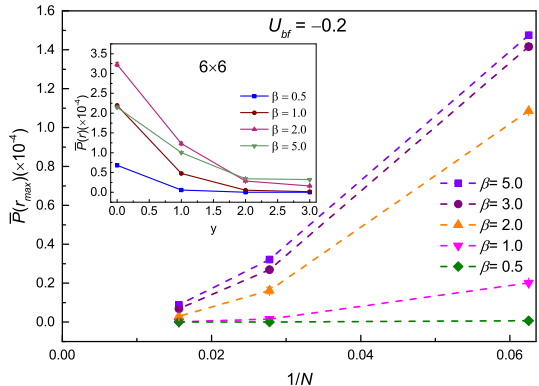}\\
  \caption{Finite size extrapolations for the off-diagonal p-wave pair correlation function $\bar P(r_{max})$ at $r=r_{max}$ along y-direction of lattice with the Bose-Fermi interaction $U_{bf}=-0.2$ at various inverse temperature $\beta=0.5,1.0,2.0,3.0,5.0$ as a function of lattice size $1/N$. Here $N =4\times4, 6\times6, 8\times8$. The inset is the p-wave pair correlation functions $\bar P(r)$ for lattice size $6\times6$ along y-direction at various inverse temperature $\beta = 0.5, 1.0, 2.0, 5.0$ with $U_{bf}=-0.2$.}\label{Fig7}
\end{figure}
The results of $p$-wave pair correlation function $\bar P(r_{max})$ have been shown in Fig. \ref{Fig7}. It is observed that the $p$-wave pair correlation function decays quickly to zero as a function of lattice size at the high temperature $\beta=2.0,1.0,0.5$, while appears a finite value at low temperature $\beta=5.0,3.0$. To verify the superfluidity of the fermion unquestionably in the thermodynamic limit, larger size and low temperature are still needed.

\subsection{\label{sec:level2} Composite pair correlation function}
In this subsection, we concentrate on the correlations between the bosons and fermions which can be characterized by the composite pair Green¡¯s function. A general definition of the Green¡¯s function is $G_{bf}(r)=\langle \Delta^\dagger (j) \Delta (i)  \rangle$ with $\Delta (i) = b_i c_i$. Due to the particle-hole transformation, the sign of $G_{bf}$ changes when the interaction $U_{bf}$ is changed to $-U_{bf}$. For the sake of clarity, we redefine the composite pair correlation as :
\begin{equation}\
\bar{G}_{bf}\left( r \right) = G_{bf}\left( r \right) - \left\langle  {b_{_j}^\dag {b_i}}  \right\rangle \left\langle  {c_{_j}^\dag {c_i}} \right\rangle
\end{equation}
By using this definition, the composite pair correlation is equal to zero when there is no interaction between the bosons and fermions. \begin{figure}
  \centering
  \includegraphics[width=8.4cm]{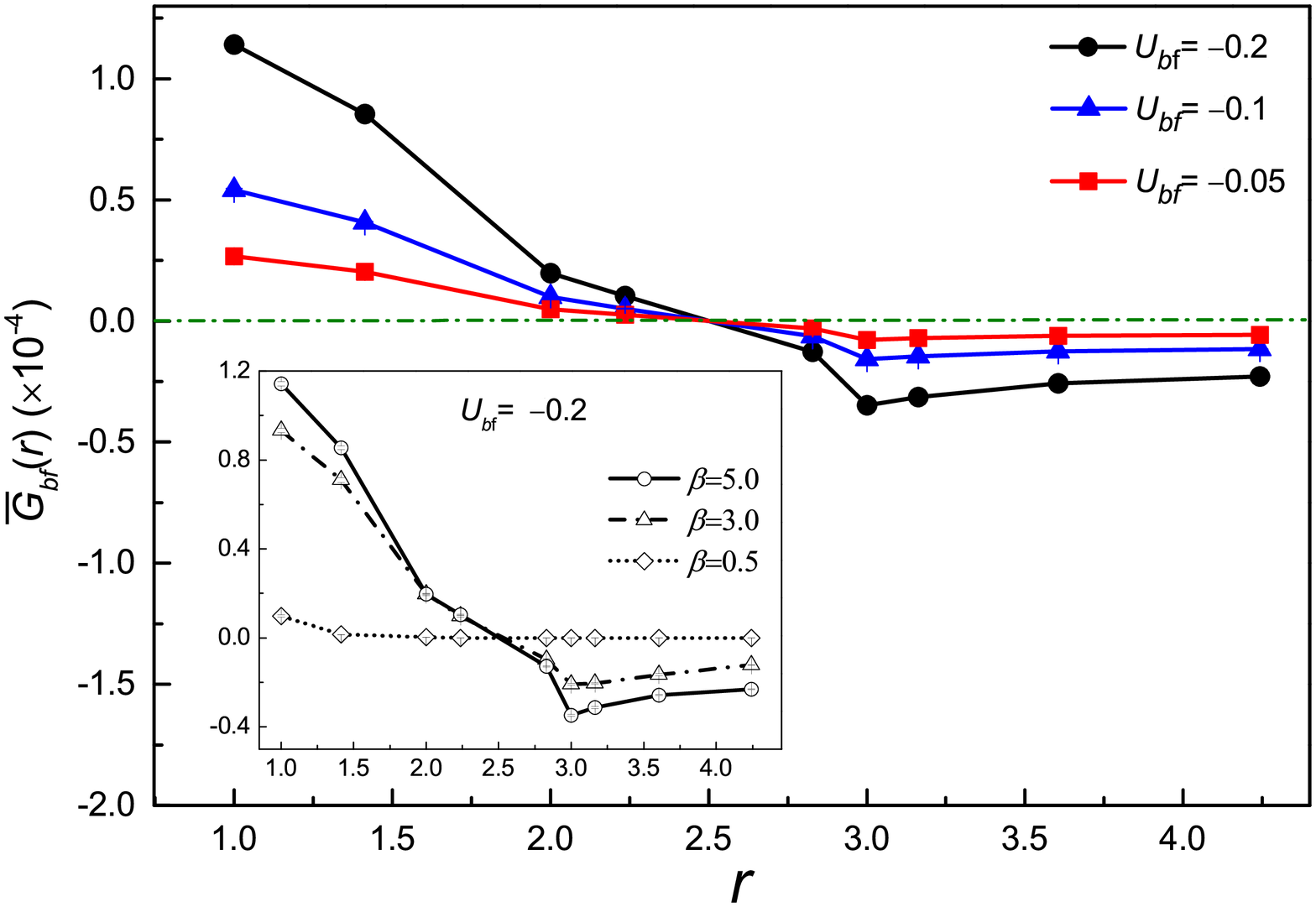}\\
  \caption{The Bose-Fermi composite pair correlation as a function of $r$ for various Bose-Fermi interactions $U_{bf}=-0.05,-0.1,-0.2$ (circle line) at inverse temperature $\beta=5.0$. The inset displays the composite pair correlation function for various inverse temperatures $\beta=5.0,3.0,0.5$ at $U_{bf}=-0.2$.}\label{Fig8}
\end{figure}
The numerical results of $\bar{G}_{bf}(r)$ is presented in Fig. \ref{Fig8}. It is obvious that the stronger interaction between bosons and fermions may result in stronger composite pair correlation. However, the evolution of $\bar{G}_{bf}(r)$  as a function of $r$ shows a non-monotonic behavior. The values of $\bar{G}_{bf}(r)$ for different $U_{bf}$ intersects and vanishes around $r=2.5$ and reaches the negative maximum around $r=3$ and then become much suppressed as $r$ further increases. As the inset displayed, the correlation becomes smeared out by increasing temperatures.

\section{\label{sec:level1} Summary}
In summary, we have studied the 2D Bose-fermi mixture on square lattice by using the DQMC method at a finite temperature in the ranges where the sign problem can be ignored. Our results show the boundary of the collapsed state at various temperatures in the attractive Bose-Hubbard model. More interestingly, when the interaction between bosons and fermions turns on, the off-diagonal long-range order of composite fermions pair shows a fairly flat finite value at a distance $r>2.5$ at low temperature, which denotes that the composite fermi pairs exist in the mixture. Meanwhile, we also have observed a slowing decay $p$-wave pair function due to the effective attractive interaction induced by the superfluid boson, which denotes that the $p$-wave superconductivity may be detected in the Bose-Fermi mixture. This super-pairing mechanism is different from the conventional BCS pairing mechanism in which effective interactions are induced by exchanging massless phonons. Moreover, to show finite size effect, finite size extrapolations have been done for $p$-wave pair correlation function at various inverse temperatures. It is obvious that a larger finite value can be observed as the temperature goes down. To verify the superfluid of the fermion unquestionably  in the thermodynamic limit, larger size system and lower temperature are needed.

$Acknowledgements.-$ We are grateful to B. M. Rubenstein, Z. B. Huang and Qi-Bo Zeng for useful discussions about some detail of the program of BF-DQMC.
This work was supported by the National Key Research and Development Program of China (Grants No. 2017YFA0304204 and No. 2016YFA0300504), the National Natural Science Foundation of China (Grants No. 11625416 and No. 11474064), and the Shanghai Municipal Government under the Grant No. 19XD1400700.

%\bibliographystyle{unsrt}%
%\bibliography{bib_file}

\end{document}